# A high resolution rovibronic molecular cross-section of MgH$^+$ molecular cation

**Huagang Xiao[a] and Tao Gao[a*]**

[a] *Institute of Atomic and Molecular physics, Sichuan University, Chengdu 610065, China*

E-mail: gaotao@scu.edu.cn

**ABSTRACT:** The high resolution rovibronic line list of MgH+ molecular cation are presented in our work. The potential energy curves are calculated by the method of multireference configuration interaction plus Davidson correction (MRCI+Q) and spin-orbit coupling (SOC) effect. Spectroscopy constants are fitted and the results are in good agreement with the experiment, ensuring the accuracy of the electronic structure. On account of potential energy curves and transition dipole moments, the Franck - Condon factors and Einstein coefficients of transition are obtained. These calculations are used to obtain an accurate partition functions and line list for the molecule. Using the data obtained from the *ab initio* calculation, the absorption cross-sections under different temperatures and pressures were simulated. Our work could provide some theoretical insights into solar and cold planet spectrum.

[*] Corresponding author: Institute of Atomic and Molecular physics, Sichuan University, Chengdu, China

# Contents

# 1. Introduction

Mg is the ninth most abundant element in the universe[1] and MgH is an important astrophysical molecule[2-4]. MgH$^+$ molecular ions may play a role in astrophysics by existing as intermediates (e.g, OH$^{+5}$ and HeH$^{+6,\ 7}$ molecular cation). As early as thirty years ago, Singh[8] have reported that MgH$^+$ is solar molecule, and its A-X transition is of great significance to the spectroscopic research of the sun and cold planets. In addition, Pearse[9] observed the emission spectrum of MgH molecule and found a extensive system of red-degraded bands between 250 nm and 340 nm, which correctly attributed it to the $^1$S - $^1$S transition of the MgH$^+$. Therefore, it is necessary for astrophysics to study MgH$^+$ molecular ions.

The MgH$^+$ molecular cation was first reported by Pearse et al. The γ -system spectrum in the Ultra- Violet spectrum of MgH molecules is the ionization of MgH$^+$ ion. In 1976, Singh and Maciel[10] et al discussed the possible existence of MgH$^+$ molecular cation in the sun. Since then, researchers have continued to discuss this topic. However, whether MgH$^+$ molecular ions exist in the solar spectrum requires the location of the line positions and dissociation energies. Therefore, it is meaningful to study the line list of MgH$^+$ molecular cation using use high precision calculations. The accuracy of the molecular electronic structure is the primary guarantee for ensuring the spectrum. In 1971, Based on the experimental data, Balfou[11] used RKR method to conduct rotational analysis of the A$^1\Sigma^+$ - X$^1\Sigma^+$ and B$^1\Pi$ - X$^1\Sigma^+$ transitions, and the partial energy level values and Franck - Condon factors of the molecule is obtained. Dalleska *et al.*[12] used Guided ion beam mass spectrometry to investigate the reaction of ground state Mg$^+$ and Sr$^+$ with molecular hydrogen and its isotopes, and determined the dissociation energy( 1.94 ± 0.06 eV) of MgH$^+$ molecular ions. In 2005, Jørgensen and Drewsen[13] used the multiconfiguration self-consistent field (MCSCF) approach to calculate the four electronic states (X$^1\Sigma$, A$^1\Sigma$, B$^1\Pi$ and C$^1\Sigma$) of the MgH$^+$ , and compared the rotational energy and dissociation energy with the experimental values of Balfou et al. It is noticed that the energy of the low level is in good agreement with the experimental value, but the energy of the high level has a certain deviation from the experiment. Four years later, Aymar *et al*[14] used different basis sets to calculate the potential energy curves and dipole moments of the $MgH^+$ molecular ion, which showed that large basis sets could more accurate. In 2013, Khemiri *et al*[15] used *ab initio* theory to extend the research of aymer et al and calculated the high excited electronic state of $MgH^+$ molecular ion.

The study of $MgH^+$ molecular ions is currently incomplete, such as the consideration of spin-orbit coupling effects. Our goal is to provide a high resolution line list of MgH+ molecular cation based on accurate electronic structures, which could increase theoretical insights for astrophysical applications.

## 2. Method

### 2.1 Electronic structure and transition property

The electronic structure and transition dipole moments are calculated using Molpro 2012 program[16]. The complete active space self-consistent[17, 18] (CASSCF) and multireference configuration interaction plus Davidson corrections[19, 20] (MRCI+Q) method are employed in the calculation. To improve accuracy, the correlation consistent polarized valence Quintuple - $\zeta$ basis set (AV5Z) is applied for Mg[21] and H[22] atoms. The diatomic molecule MgH+ belongs to $C_{\infty v}$ group，However, the $C_{2v}$ subgroup is adopted limited by the Molpro code. The $C_{2v}$ group contain four irreducible representations ($a_1$, $b_1$, $b_2$, $a_2$), which is related to the electronic state and orbital of the molecule(e.g, $\Sigma^+ = a_1$, $\Pi = b_1 + b_2$, $\Sigma^- = a_2$). In our calculation, the 1s2s2p orbital electron of Mg atom are placed into frozen space and the 3s3p orbital of and 1s orbital of H are putted into active space. Considering the calculation of excited states, electrons are excited from lower orbits to higher orbits. The 4s4p orbital of Mg atom also considered into active space. Hence, the total spaces of $MgH^+$ molecular cation are expressed as ($8a_1$, $3b_1$, $3b_2$, $0a_2$), which could ensured the accurate calculation of six $\Lambda$ - S states and eleven $\Omega$ states. Using the LEVEL program[23], the spectroscopic constants of electronic states are obtained.

Combining the potential energy curves and the transition dipole moments, the transition properties of the molecular electronic state could be obtained. The Franck - Condon factor $f_{v'v''}$ and Einstein spontaneous emission coefficient $A_{v'v''}$ are described by the follows:

$$f_{v'v''} = |\int \Psi_{v'J'}\Psi_{v''j''}dr|^2 \quad (1)$$

$$A_{v'v''} = 3.1361891\times10^{-7}\frac{S(J',J'')}{2J'+1}v^3\langle\Psi_{v',J'}|M(r)|\Psi_{v'',J''}\rangle \quad (2)$$

Where $v'J'$ and $v''J''$ represents upper level and lower level, $\psi_{v'J'}$ and $\psi_{v''J''}$ is wave functions for upper and lower levels, M(r) is transition dipole moments and S(J', J'') is the Hönl-London rotational intensity factor of levels.

**2.2 Spectral simulation**

Here, a brief description of the molecular absorption spectrum calculation is shown below. First, solving the radial Schrödinger equation of the electronic state, the vibrational level and rotational level of the molecule is obtained[24].

$$-\frac{h^2}{2\mu}\frac{d^2\Psi_{vJ}(r)}{dr^2} + V_J(r)\Psi_{vJ}(r) = E_{vJ}\Psi_{vJ}(r) \quad (3)$$

$$E_{vJ} = G(v) + B_v[J(J+1-\Omega^2)] - D_v[J(J+1-\Omega^2)]^2... \quad (4)$$

Where $E_{vJ}$ is the energy of ro-vibtational level, G(v) is energy of vibrational level, the $B_v$ and $D_v$ are centrifugal distortion constants.

In view of specific temperature, the temperature - dependent partition function Q (T) of the ro - vibrational energy level of the molecule is[25]:

$$Q_{evJ} = \sum_e\sum_{v=0}^{v_{max}}\sum_{J=\Lambda}^{J_{max}} g_J \times \exp\left\{-\frac{hc}{kT}[T_e + G(v) + F_v(J) - E_0]\right\} \quad (5)$$

Where Te is vertical transition energy of electronic states, G(v) is vibrational level energy, F(v) is rotational level energy, E$_0$ is reference energy. k, h and c represent Boltzmann's constant coefficient, Planck's coefficient and the speed of light, respectively. g$_J$ represents the degeneracy of the molecular level, which has to do with total spin and nuclear spin of the molecule. For MgH$^+$ molecular cation, the nuclear spin of Mg and H atoms are 0 and 1/2, respectively. Hence, with regard to the all ro-vibrational levels , the total molecular spin is 1/2 and the spin multiplicity g$_N$ is 2. Integrated intensity Sul of each electronic vibration rotational transition spectral line:

$$S_{ul} = \frac{1}{8\pi c}(E_u - E_l)^{-2} \frac{g_u e^{-\left(\frac{hc}{kT}\right)E_l}\left[1-e^{-\left(\frac{hc}{kT}\right)(E_u-E_l)}\right]}{Q(T)} A_{ul} \qquad (6)$$

The absorption cross section $\sigma_{ul}$ is closely related to the intensity $S_{ul}$,

$$S_{ul} = \int_{-\infty}^{\infty} \sigma_{ul}(v) dv \qquad (7)$$

In above formula, $g_u$ represents the degeneracy of the higher level, which is related to the nuclear spin. E$_u$ and E$_l$ are represent the energy of higher and lower levels, respectively.

## 3. Results

### 3.1 Electronic structure and ro-vibrational energy levels

Molecular structure and properties could be described by potential energy curves (PECs). The PECs of six $\Lambda$ - S states were depicted in Fig. 1(a), which including four bond states(X$^1\Sigma^+$, A$^1\Sigma^+$, B$^1\Pi$ and b$^3\Pi$) and two repulsed states(a$^3\Sigma^+$ and c$^3\Sigma^+$). With the spin orbital couple effect(SOC), six $\Lambda$ - S states splitted into eleven $\Omega$ states and the result are shown in Fig. 1(b). Meanwhile, the Mg$^+$ ($^2$P$_u$) splitted two energy levels ($^2$P$_{1/2}$ and $^2$P$_{3/2}$) with an energy level difference of 69.2 cm$^{-1}$, which is good agreement with the experiential value[26] (91.5 cm$^{-1}$).

Table 1. The dissociation relationships of $\Omega$ states

| Atomic states | $\Omega$ states | Cal. energy (cm$^{-1}$) | Expt. Energy[26] (cm$^{-1}$) |
|---|---|---|---|
| Mg$^+$ ($^2$S$_{1/2}$) + H ($^2$S$_{1/2}$) | X$^1\Sigma^+_{0+}$, a$^3\Sigma^+_{0-}$, a$^3\Sigma^+_1$ | 0 | 0 |
| Mg$^+$ ($^2$P$_{1/2}$) + H ($^2$S$_{1/2}$) | A$^1\Sigma^+_{0+}$, b$^3\Pi_{0-}$, b$^3\Pi_1$ | 34482.5 | 35669.3 |
| Mg$^+$ ($^2$P$_{3/2}$) + H ($^2$S$_{1/2}$) | b$^3\Pi_{0+}$, b$^3\Pi_2$, B$^1\Pi_1$, c$^3\Sigma^+_{0-}$, c$^3\Sigma^+_1$ | 34552.1 | 35760.8 |

The spectroscopic parameters of those electronic states were fitted and were listed in Table 2. The $X^1\Sigma^+$, $A^1\Sigma^+$ and $B^1\Pi$ bond states are no splitting and it is notable that spectroscopic parameters of those states are changed little between spin-free and SOC effect. The ground state $X^1\Sigma^+$ possess a deep potential well of 16955.9 cm$^{-1}$, which is closer to experimental data[27] (16937 cm$^{-1}$) than other theoretical values. In addition, it found that the equilibrium length $R_e$, harmonic constant $\omega_e$ and rotational constant $B_e$ of $X^1\Sigma^+$ state is better match the expermental data than other theoretical calculation. And this phenomenon also occur at the excited states $A^1\Sigma^+$ and $B^1\Pi$. The electronic state $b^3\Pi$ lying 46561.3 cm$^{-1}$ above ground state $X^1\Sigma^+$ with a shallow potential well 4882.2 cm$^{-1}$ (0.605 eV). The equilibrium length $R_e$ of $b^3\Pi$ is 2.030 Å, which is close to the data ($R_e$ = 2.032 Å) calculated by khemiri $et\ al$ using effective core potential (ECP) + core polarization (CPP) + full configuration interaction (FCI) approach. Aymar[14] $et\ al$ also employed ECP + CPP + FCI method to calculate the electronic properties of MgH$^+$ molecular cation. The spectroscopic parameters of $b^3\Pi$ are far from what our calculated, however, the $D_e$, $R_e$ and $\omega_e$ of $c^3\Sigma^+$ state reported by Aymar $et\ al$ are agreement with the values of $b^3\Pi$, which may be an wrote error for them. With the SOC effect, $b^3\Pi_{0^-}$, $b^3\Pi_{0^+}$, $b^3\Pi_1$ and $b^3\Pi_2$ are splitted from $b^3\Pi$ state. Notably, the split strength of $b^3\Pi$ is inconspicuous and only 42.8 cm$^{-1}$ energy gap generated between the $b^3\Pi_{0^-}$ and $b^3\Pi_2$ electronic states.

Comparing with the other theoretical calculation and expiential data, the ability of our calculation to accurately predict the electronic structure and spectroscopic constants and indicate that our present results are most sophisticated and accurate work done for the low-lying states of MgH$^+$ molecular cation.

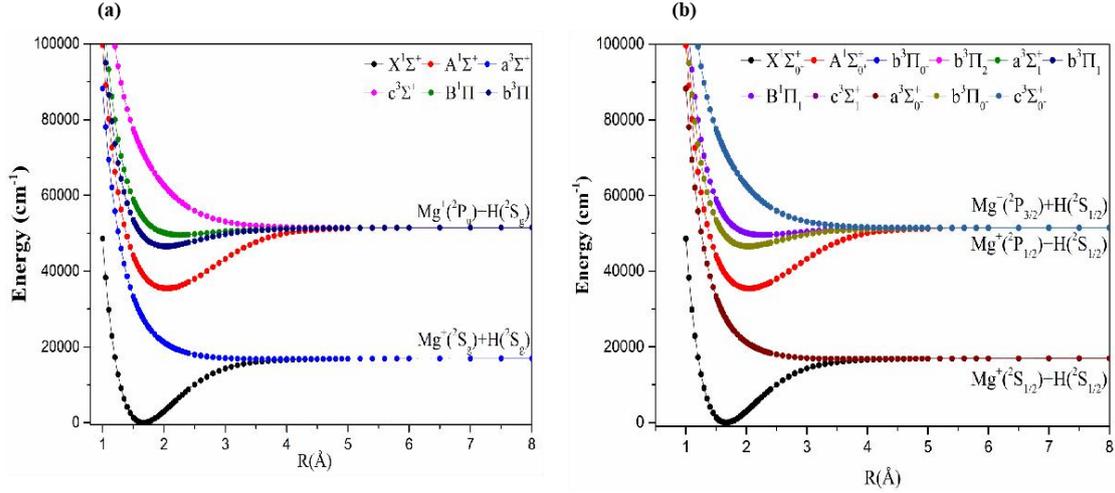

**Fig. 1** The potential energy curves of MgH$^+$ molecular cation. (a) the Λ - S states for two channels, (b) the Ω states for three channels with spin-orbital coupling effect.

Table 2. Spectroscopic parameters of the bond states for MgH$^+$ molecular cation from MRCI+Q+SOC calculations compared to other Theoretical and Experimental results

| | $D_e$/cm$^{-1}$ | $T_e$/cm$^{-1}$ | $R_e$/Å | $\omega_e$/cm$^{-1}$ | $B_e$/cm$^{-1}$ | Reference |
|---|---|---|---|---|---|---|
| X$^1\Sigma^+$ | 16955.9 | 0 | 1.665 | 1677.4 | 6.2685 | MRCI+Q |
| X$^1\Sigma^+_{0+}$ | 16955.9 | 0 | 1.665 | 1684.1 | 6.2895 | MRCI+Q+SOC |
| | 15650±500 | | | | | Expt.[12] |
| | 16937 | 0 | 1.649 | 1695.3 | 6.411 | Expt.[27] |
| | 16501 | 0 | 1.635 | 1599 | | Thery[14] |
| | 14755 | 0 | 1.667 | 1542 | | Thery[15] |
| A$^1\Sigma^+$ | 15837.4 | 35433.5 | 2.050 | 1109.6 | 4.1645 | MRCI+Q |
| A$^1\Sigma^+_{0+}$ | 15832.4 | 35433.4 | 2.050 | 1115.6 | 4.1641 | MRCI+Q+SOC |
| | | 35905 | 2.006 | 1132.7 | 4.330 | Expt.[27] |
| | 16656 | | 2.005 | 1125 | | Thery[14] |
| | 15666 | 37689.1 | 2.001 | 1129 | | Thery[15] |
| B$^1\Pi$ | 1874.1 | 49593.0 | 2.300 | 557.2 | 3.3788 | MRCI+Q |
| B$^1\Pi_1$ | 1892.9 | 49593.2 | 2.300 | 555.7 | 3.3788 | MRCI+Q+SOC |
| | 1943 | | 2.27 | 588 | 3.39 | Thery[14] |
| | 1769 | 53009.4 | 2.28 | 549 | | Thery[15] |
| b$^3\Pi$ | 4882.2 | 46561.3 | 2.030 | 951.4 | 4.2223 | MRCI+Q |
| | 5057 | | 2.005 | 915 | | Thery[14] ($c^3\Sigma^+$) |
| | 79 | | 4.905 | 70 | | Thery[14] ($b^3\Pi$) |
| | 4654 | | 2.032 | 83 | | Thery [15] |
| $b^3\Pi_0^-$ | 4868.5 | 46539.8 | 2.030 | 952.6 | 4.2217 | MRCI+Q+SOC |
| $b^3\Pi_0^+$ | 4960.7 | 46539.9 | 2.030 | 921.7 | 4.2217 | MRCI+Q+SOC |
| $b^3\Pi_1$ | 4850.4 | 46561.1 | 2.030 | 956.1 | 4.2215 | MRCI+Q+SOC |
| $b^3\Pi_2$ | 4883.9 | 46582.6 | 2.030 | 948.3 | 4.2214 | MRCI+Q+SOC |

Fine molecular structure could bring out accurate energy level. Solving the one-dimensional Schrödinger equation, the vibrational levels and rotational levels of the ground and excited states are obtained. The radiative transition of $\Omega = 0 - 0$ is strongest, therefore, our interest mainly focus on 0 - 0 transition and the parameters of energy level for $X^1\Sigma^+_{0+}$, $A^1\Sigma^+_{0+}$ and $b^3\Pi_0^+$ states are listed in Table. 3. The ground state $X^1\Sigma^+_{0+}$ own 17 vibrational levels and the rotational level $J \leq 69$. The excited state $A^1\Sigma^+_{0+}$ also presented deep well, corresponding to a maximum vibrational quantum number $\upsilon_{max}$ of 19 and a maximum rational quantum number $J_{max}$ of 90. Notably, the gap of vibrational energy levels for different electronic state are not alike, which is due to the distinction of harmonic constant $\omega_e$. Partial vibrational level energy are showed in Tables 3 and All rovibronic level energies are attached to the Support material. Compared with experimental data, it could found that the energy value is good agreement with experiment.

Table. 3 the vibrational quantum parameters of energy level for MgH$^+$ molecular cation

| Electronic states | vibrational quantum number $\upsilon$ | $E_{cal}$ | $E_{expt.}$[11] |
|---|---|---|---|
| $X^1\Sigma^+_{0+}$ | 0 | 841.318 | 841.57 |
| | 1 | 2473.976 | 2476.13 |
| | ... | ... | ... |
| | 17 | 16955.954 | 16937 |
| $A^1\Sigma^+_{0+}$ | 0 | 547.485 | 565.85 |
|   | 1 | 1627.996 | 1684.81 |
|   | ... | ... | ... |
|   | 19 | 15837.419 | 15920 |
| $b^3\Pi_0^+$ | 0 | 443.317 |   |
|   | 1 | 1277.759 |   |
|   | ... | ... |   |
|   | 10 | 4960.709 |   |

Notes.

$\upsilon$: vibrational quantum number

$J$: rotational quantum number

$E$: energy of levels/cm$^{-1}$

$g_N$: the degenerate of levels

### 3.2 Transition properties

It is particularly important to understand the transition dipole moments (TDMs) when investigating the transition strength between the various electronic states. Since the $b^3\Pi \rightarrow X^1\Sigma^+$ band is forbidden, the SOC effect is considered in our calculation. The DMs of $X^1\Sigma^+_{0+} \rightarrow X^1\Sigma^+_{0+}$, $A^1\Sigma^+_{0+} \rightarrow X^1\Sigma^+_{0+}$ and $b^3\Pi_0^+ \rightarrow X^1\Sigma^+_{0+}$ transitions are shown in Fig. 2. The DM of $X^1\Sigma^+_{0+} \rightarrow X^1\Sigma^+_{0+}$ transition is permanent dipole moments of the ground state $X^1\Sigma^+_{0+}$ and the max value is 1.472 a.u (3.74 Debye) when the equilibrium bond length R is 1.4 Å. The transition dipole moments (TDMs) of $A^1\Sigma^+_{0+} \rightarrow X^1\Sigma^+_{0+}$ first increased to a maximum value of 2.1 a.u (5.34 Debye) and then slowly decreased. However, compared with the first excited state $A^1\Sigma^+_{0+}$, the TDM values of $b^3\Pi_0^+ \rightarrow X^1\Sigma^+_{0+}$ transition always near to zero in the Franck - Condon region.

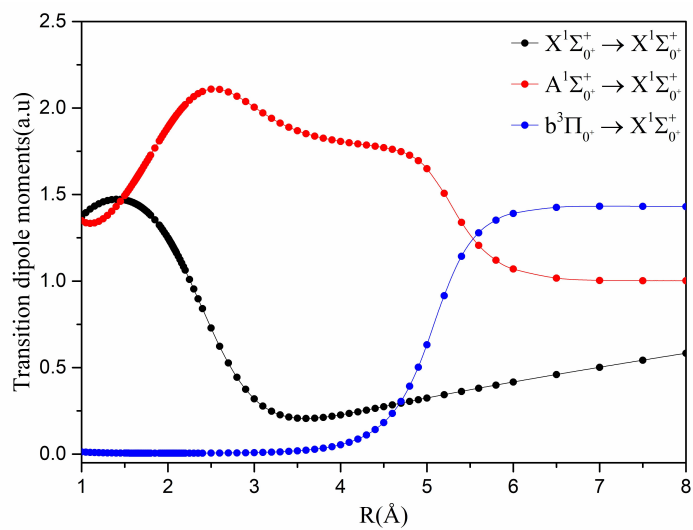

**Fig. 2** Transition dipole moments between $X^1\Sigma^+_{0+}$, $A^1\Sigma^+_{0+}$ and $b^3\Pi_{0^+}$ states of MgH$^+$ molecular cation.

Franck - Condon factors and Einstein coefficients are importance characterization parameters for our investigation of transition properties. Relaying on the potential energy curves and dipole moments, the Franck - Condon factors $f_{ul}$ and Einstein coefficients $A_{ul}$ of $A^1\Sigma^+_{0+}$ - $X^1\Sigma^+_{0+}$ and $b^3\Pi_0^+$ - $X^1\Sigma^+_{0+}$ transitions are obtained using LEVEL program and the results are listed in Table. 4. It is notice that the $f_{ul}$ not show a diagonal trend due to the difference in equilibrium bond length $R$ between the excited states ($A^1\Sigma^+_{0+}$ and $b^3\Pi_0^+$) and the ground state ($X^1\Sigma^+_{0+}$). In 1971, Balfour[11] evaluated the FCFs of $A^1\Sigma^+$ - $X^1\Sigma^+$ transition using RKR method, However, its value is slightly different from our calculation. The reason is because of the error in the equilibrium bond length $R_e$ our calculated (our value: $R_e$ ($X^1\Sigma^+_{0+}$) is 1.665 Å, Balfour's data: $R_e$ ($X^1\Sigma^+_{0+}$) is 1.652 Å). It could found that the Einstein coefficients $A_{ul}$ of $A^1\Sigma^+_{0+}$ - $X^1\Sigma^+_{0+}$ transition is larger and the values reach $10^7$ s$^{-1}$, which is owing to the large TDMs. The excited state $b^3\Pi$ is lying 46561.3 cm$^{-1}$ above ground state and the $b^3\Pi$ - $X^1\Sigma^+$ bond is forbidden transition. Considering the spin - orbital coupling (SOC) effect, $b^3\Pi_0^+$ - $X^1\Sigma^+_{0+}$ transition is obtained in our calculation. Similar to the first excited state $A^1\Sigma^+_{0+}$, the max Franck - Condon factor is $f_{02}$ (0.2976). The TDMs of $b^3\Pi_0^+$ - $X^1\Sigma^+_{0+}$ transition is much smaller than $A^1\Sigma^+_{0+}$ - $X^1\Sigma^+_{0+}$ transition, which bring out minor $A_{ul}$ of $b^3\Pi_0^+$ - $X^1\Sigma^+_{0+}$ transition.

Table. 4 The Franck - Condon factors $f_{ul}$ and Einstein coefficients $A_{ul}$ (s$^{-1}$) of two transitions for MgH$^+$ molecular cation.

| Transition | | $v'=0$ | $v'=1$ | $v'=2$ | $v'=3$ |
|---|---|---|---|---|---|
| A$^1\Sigma^+_{0+}$ - X$^1\Sigma^+_{0+}$ | $v''=0$ | 0.0666 | 0.1531 | 0.1975 | 0.1876 |
| | | 0.091[a] | 0.185[a] | 0.214[a] | 0.185[a] |
| | | 1.84E+7 | 4.40E+7 | 5.88E+7 | 5.77E+7 |
| | $v''=1$ | 0.2277 | 0.1946 | 0.0530 | 0 |
| | | 0.273[a] | 0.175[a] | 0.025[a] | 0.007[a] |
| | | 5.90E+7 | 5.24E+7 | 1.47E+7 | 8650.17 |
| | $v''=2$ | 0.3268 | 0.0194 | 0.0520 | 0.1144 |
| | | 0.337[a] | 0.002[a] | 0.093[a] | 0.114[a] |
| | | 7.92E+7 | 4.86E+6 | 1.39E+7 | 3.15E+7 |
| | $v''=3$ | 0.2493 | 0.0811 | 0.1275 | 0.0049 |
| | | 0.216[a] | 0.149[a] | 0.097[a] | 0.002[a] |
| | | 5.64E+7 | 1.94E+7 | 3.18E+7 | 1.22E+6 |
| b$^3\Pi_{0+}$ - X$^1\Sigma^+_{0+}$ | $v''=0$ | 0.0806 | 0.1510 | 0.1685 | 0.1475 |
| | | 58.30 | 221.79 | 363.16 | 382.12 |
| | $v''=1$ | 0.2296 | 0.1701 | 0.0501 | 0.0021 |
| | | 8.53 | 123.57 | 172.55 | 78.19 |

| | | | | |
|---|---|---|---|---|
| v″ = 2 | 0.2976 | 0.0213 | 0.0285 | 0.0806 |
| | 163.77 | 0.21 | 19.08 | 3.05 |
| v″ = 3 | 0.2292 | 0.0449 | 0.1230 | 0.0364 |
| | 582.62 | 3.16 | 24.29 | 11.26 |

[a] the experiential value from reference [11]

### 3.3 Temperature-dependent partition function

The values of partition function were calculated in 50 K - 30000 K range, whose temperature could span from low temperature, medium to high temperature to high temperature. Depending on the precise ro - vibrational level energy, the total partition function and proportion of each states are depicted in Fig. 3 and the data are listed in Support material. With the temperature increasing, the partition function Q also gradually increases. However, it could found that the molecular partition functions were almost contributed by the ground state $X^1\Sigma^+_{0+}$ below 8000 K. When the temperature continues to increase, the proportion partition functions of the excited state gradually reveal. For example, the partition function of the ground state $X^1\Sigma^+_{0+}$ and first excited state $A^1\Sigma^+_{0+}$ at temperature T = 15000 K calculated are 20874.3 and 1411.2, whose proportion are 89.9% and 6.1%. Moreover, it found that within a temperature of 30000 K, the sum of the partition functions of the molecule is 1, indicating that it is completely sufficient for us to calculate the electronic state of $MgH^+$ molecular cation.

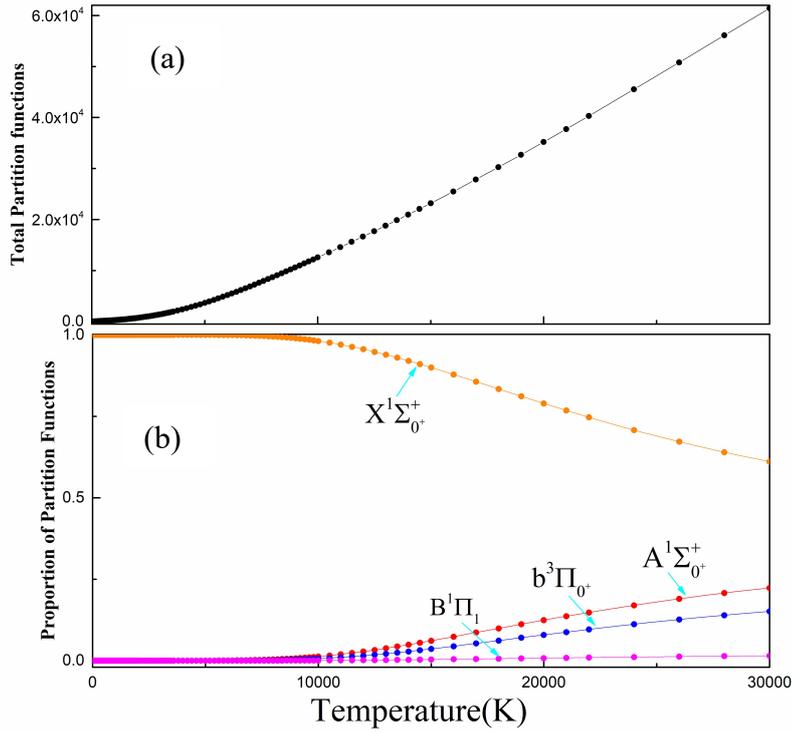

**Fig. 3** the partition function of MgH+ molecular cation: (a) the total partition function, (b) the proportion of partition function for each electronic states.

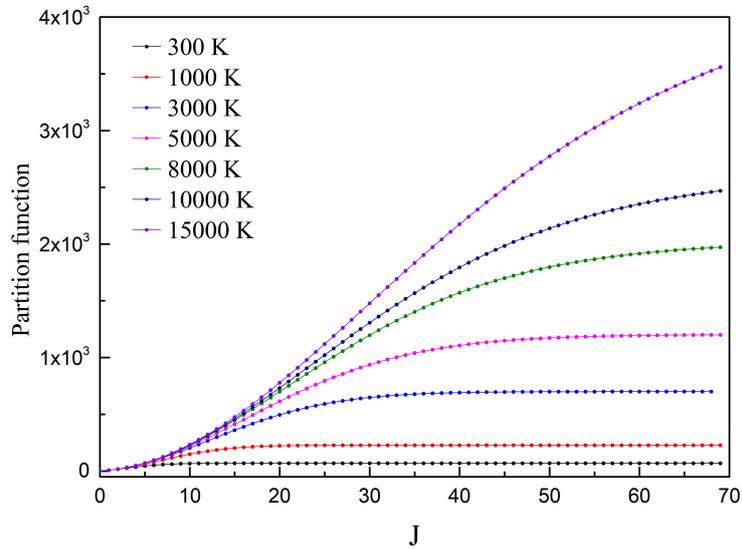

**Fig. 4** the partition function Q(T) of MgH$^+$ molecular cation with respect to the rotational quantum number $J$ ($v = 0$) at different temperature.

It clear found that the ground state plays a crucial role under 30000 K. Hence, the partition function of the rotational quantum number $J$ ($v = 0$) for ground state $X^1\Sigma^+_{0+}$ at different temperatures T is shown in Fig. 4. As the temperature increases, the convergence of the partition with J is slower. When the temperature reaches above 8000 K, the values continue to grow, indicating that more energy levels are involved.

### 3.4 Simulated spectra

A ro - vibrational line lists of MgH+ molecular cation are computed using Duo program[28]. As before mentioned, the Einstein coefficients of $b^3\Pi_{0+}$ - $X^1\Sigma^+_{0+}$ transition is smaller, hence, Our attention is mainly focused on the $X^1\Sigma^+_{0+}$ - $X^1\Sigma^+_{0+}$ transition and the $A^1\Sigma^+_{0+}$ - $X^1\Sigma^+_{0+}$ transition. Combined the *.states* and *.trans* files of the Duo program and using the EXOCROSS code[29], the simulations spectrum of MgH+ molecular cation is obtained.

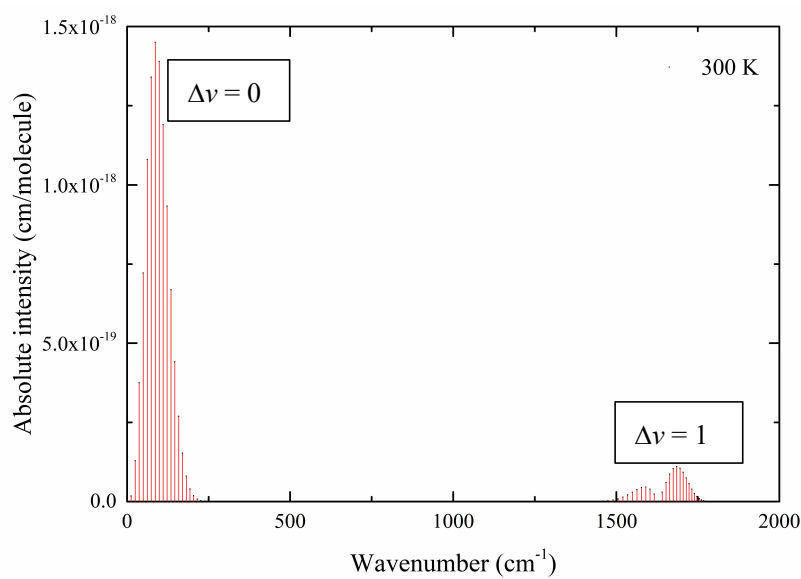

**Fig. 5** the absolute absorption line intensities of $X^1\Sigma^+_{0+}$ - $X^1\Sigma^+_{0+}$ transition for MgH$^+$ molecular cation at T = 300 K.

A MgH$^+$ stick spectrum of $X^1\Sigma^+_{0+}$ - $X^1\Sigma^+_{0+}$ transition at T = 300 K is plotted in Fig. 5, which contain the two terms ($\Delta v$ = 0 and 1). Obviously, the $\Delta v$ = 0 term in a pure rotational spectrum is stronger than the $\Delta v$ = 1. In addition, It could found that the P and R branches are clearly visible in $\Delta v$ = 1 term.

In order to make the research more systematic, the controlled variable method is adopted to investigate the molecular spectrum. First, The absorption cross sections of MgH$^+$ molecular ions under different pressures at temperature T = 300 K are shown in Fig. 6. Under the same pressure(eg, 1.0 bar), the intensity of $X^1\Sigma^+_{0+}$ - $X^1\Sigma^+_{0+}$ is 10$^{-3}$ orders of magnitude smaller than the intensity of $A^1\Sigma^+_{0+}$ - $X^1\Sigma^+_{0+}$. The max value of cross-section of $X^1\Sigma^+_{0+}$ - $X^1\Sigma^+_{0+}$ transition is $2.79 \times 10^{-19}$ cm$^2$/molecule, which represents the transition of 0 - 0 band. However, for the $A^1\Sigma^+_{0+}$ - $X^1\Sigma^+_{0+}$ transition, the max cross-section is $2.45 \times 10^{-16}$ cm$^2$/molecule, which is generated by the 3 - 0 band. Furthermore, with the temperature remains constant and pressure increases, the peak shape of the spectrum remains unchanged, but the intensity generally decreases. And then, holding the pressure constant, the effect of temperature on the spectrum is explored and the results are presented in Fig. 7. As the temperature increases, there are more levels contribute to the absorption spectrum, which making its spectral distribution wider. Notably, with the temperature increases, the transition spectrum undergoes a shift.

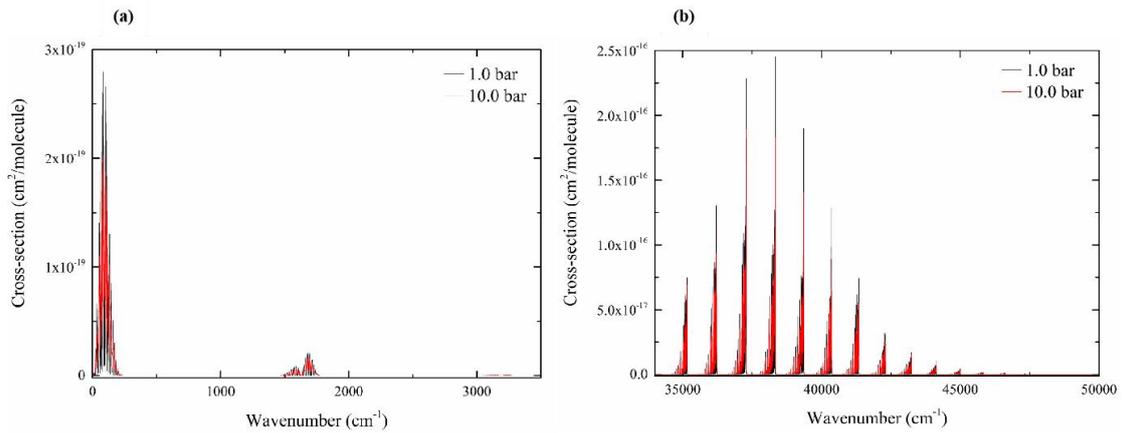

**Fig. 6** Absorption cross-sections of MgH$^+$ molecular cation under different pressures at T = 300 K.(a: $X^1\Sigma^+_{0+}$ - $X^1\Sigma^+_{0+}$ transition, b: $A^1\Sigma^+_{0+}$ - $X^1\Sigma^+_{0+}$ transition)

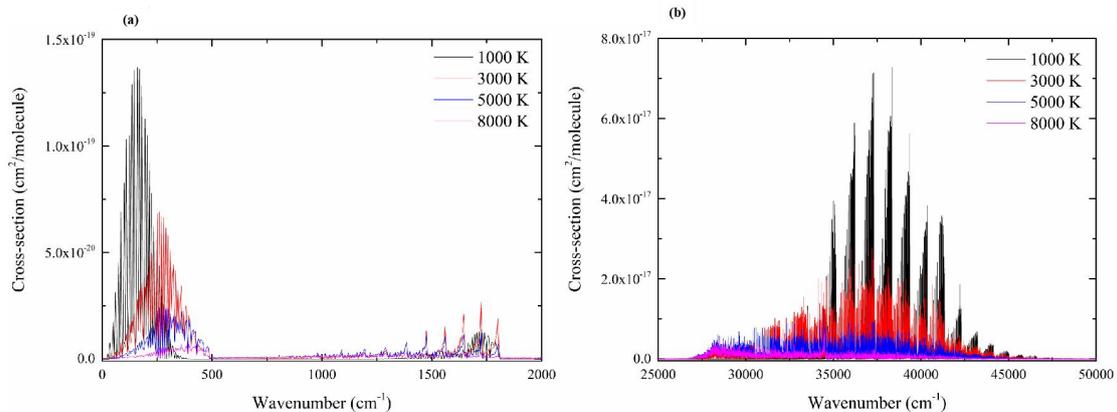

**Fig. 7** the spectrum of MgH$^+$ molecular cation in various temperature at P = 1.0 bar. (a: $X^1\Sigma^+_{0+}$ -

$X^1\Sigma^+_{0+}$ transition, b: $A^1\Sigma^+_{0+}$ - $X^1\Sigma^+_{0+}$ transition)

## 4. Calculation

Here, a detailed and systematic *ab initio* spectroscopic investigation of the $MgH^+$ molecular cation is presented for the first time. It is crucial for spectroscopic studies of the Sun and cool stars. The potential energy curves are constructed using multireference configuration interaction and Davidson correction (MRCI+Q) method. Spin - orbital coupling effect is considered in our calculation. The spectroscopic constants we obtained are in well agreement with the experiments, and the errors are all around 1%, which could ensure the accuracy of the ro - vibrational levels. Using potential energy curves and transition dipole moments, The line list of the $MgH^+$ molecular cation was calculated, and the effects of different temperatures and pressures on the high-temperature spectrum were explored. The present work was conducted for applications in astrophysics.

## Acknowledgment.

Thanks to Key Laboratory of High Energy Density Physics and Technology of Ministry of Education for providing the computing resources for this work.